\documentclass[twocolumn,aps,pra]{revtex4-2}
\usepackage{bm,bbm}
\usepackage{amsfonts,amsmath,amssymb,latexsym}
\usepackage{graphicx}
\usepackage{mathtools}
\usepackage{xcolor}
\usepackage{hyperref}
\usepackage{physics}
\begin{document} 
\title{Magnetic tuning of quantum backflow: Schr\"odinger vs Pauli system}
\author{Tomasz Rado\.zycki}
\email{t.radozycki@uksw.edu.pl}
\affiliation{Faculty of Mathematics and Natural Sciences, College of Sciences, Institute of Physical Sciences, Cardinal Stefan Wyszynski University in Warsaw, W\'oycickiego 1/3, 01-938 Warsaw, Poland} 

\begin{abstract}
Probability backflow is investigated for charged spin-$1/2$ particles in a uniform magnetic field within the lowest-Landau-level approximation. While single-mode Landau states exhibit only weak and non-robust backflow, it is shown that two-mode interference gives rise to negative probability flux in both the Schr\"odinger and Pauli formulations. In the Pauli case, orbital and spin contributions to the probability current are explicitly separated, revealing an additional interference mechanism. The external magnetic field is found to act as a tunable control parameter, simultaneously controlling the spatial overlap of orbitals and the spin dynamics. In the two-mode regime, a resonance between orbital mismatch and spin precession leads to an enhancement of backflow. Finite spatial resolution is shown not to suppress the effect within realistic coarse-graining. These results establish magnetic-field-controlled quantum backflow in Landau systems as an accessible interference phenomenon.
\end{abstract}

\maketitle

\section{Introduction}\label{intro}

Probability backflow is one of the most striking features of quantum mechanics, in which a state composed exclusively of positive momentum components can nevertheless exhibit locally negative probability current. Since its original formulation by Bracken and Melloy \cite{brack}, the phenomenon has attracted sustained interest from both foundational and operational perspectives. Subsequent studies have clarified the mathematical structure \cite{penz,eveson,yearsley,few} and its close connection to time-of-arrival theory, as developed in earlier works \cite{al1,al2,al3,kij,werner,muga,toa}. Extensions to relativistic quantum mechanics have also been investigated for Dirac and related wave equations \cite{mell,asfa,ibba}. Further aspects of the physical interpretation of quantum backflow, together with optical analogues, were discussed by Berry \cite{berry}.  Classical-wave analogues and experimental approaches inspired by quantum backflow have been reported in a variety of settings \cite{saari,kot1,kot2,elie,daniel,zhang}. However, an unambiguous experimental observation of genuine quantum probability backflow remains an open challenge \cite{real}.

In the standard Schr\"odinger picture, quantum backflow is associated with interference between momentum components. For spin-$1/2$ particles described by the Pauli equation, the probability current contains an additional spin-dependent contribution, whose hydrodynamic interpretation was introduced by Takabayasi and further developed in subsequent spin-transport approaches to quantum mechanics \cite{takabayasi,holland,shi}. The current thus decomposes into orbital and spin parts, the latter being governed by spatial variations of the spin density and capable of redistributing the local probability current even in the absence of classical forces. Recently, the interplay between these orbital and spin contributions was investigated for two-mode Pauli states, revealing their role in quantum backflow \cite{trpauli}.

External electromagnetic fields provide a natural setting in which orbital and spin degrees of freedom are simultaneously influenced. In particular, the Landau problem for charged particles in a uniform magnetic field leads to degenerate Landau levels, reflecting translational invariance and the associated guiding-center degree of freedom \cite{ll,jl}. This degeneracy allows for nontrivial superpositions of states, corresponding to different guiding centers and longitudinal momenta, and thereby enables interference effects despite the presence of strong magnetic confinement. The magnetic length sets the characteristic spatial scale governing cyclotron localization and wavefunction overlap, while in the Pauli framework the magnetic field additionally induces spin precession, modifying the structure of the probability current and the resulting transport properties \cite{fw,sakurai,ll}.

As noted above, in free space, backflow is defined for states with support on positive longitudinal momenta, ensuring a fixed propagation direction. For a uniform magnetic field parallel to the $z$-axis, translational invariance along $z$ preserves $p_z$ as a good quantum number, which can still be restricted to $p_z>0$. Therefore, in this work, backflow is understood as negative probability current along the field direction under this constraint, while the transverse dependence is encoded in the degenerate Landau levels. This definition is not specific to the chosen gauge (\ref{ceca}) but follows from the fact that the magnetic field selects a distinguished spatial direction. Choosing the field along the $z$ axis is merely a coordinate convention; for an arbitrary uniform field, the definition applies to the current component parallel to the field.

Despite extensive studies of backflow in free systems and of Landau quantization, their interplay remains comparatively unexplored. Magnetic-field effects on quantum backflow have been considered in several related settings, including azimuthal motion of electrons in a constant magnetic field \cite{strange} and quantum particles in ring geometries \cite{ring}. More recently, two-dimensional backflow in the presence of magnetic flux has also been investigated, revealing the role of spatial dimensionality and state degeneracy \cite{barbi}. However, these studies do not address the interplay between Landau localization, spin dynamics, and quantum interference. It remains unclear how magnetic confinement modifies the spatial structure of interfering modes, how spin contributions affect the backflow signal, and whether an external magnetic field can serve as a genuine control parameter for this phenomenon.

The Pauli framework provides a natural setting for addressing these questions. In this work, probability backflow is analyzed for charged spin-$1/2$ particles in Landau states within the lowest-Landau-level (LLL) approximation. Unlike the recent analysis of spin-induced backflow in free-space plane-wave Pauli states \cite{trpauli}, the effects of Landau quantization and an external magnetic field are considered. It is shown that weak backflow can arise already in single-mode states. However, its magnitude remains small and vanishes in the weak-field limit, where the region of negative flux is pushed to large distances. Thus, the magnetic field is essential for making the effect physically accessible and, in the two-mode regime, serves as genuine tunable control parameter through the spatial overlap of Landau orbitals and Zeeman spin dynamics. Their interplay leads to a matching condition between the orbital interference and spin-precession frequencies, resulting in a pronounced enhancement of backflow.

Beyond its fundamental interest, controllable quantum backflow may be relevant to quantum technologies. Tuning interference-induced probability flow with an external magnetic field could be useful for quantum state engineering or coherent transport protocols, although such applications lie beyond the scope of this work.

Finally, robustness under finite spatial resolution is examined within the framework of coarse-grained flux measurements \cite{yearsley,muga}. The backflow is found to persist for detector resolutions set by the magnetic length and the interference scale.

The paper is organized as follows. In Sec.~\ref{revis}, the probability currents in the Schr\"dinger and Pauli formalisms are reviewed together with the structure of Landau eigenstates. Section \ref{smb} analyzes single-mode behavior, while Sec.~\ref{tmqi} addresses two-mode interference and magnetic-field control of backflow. Finite-resolution effects and robustness are discussed in Sec.~\ref{res}.
  
\section{Revisiting scalar and spinor states in a uniform magnetic field}\label{revis}

\subsection{Scalar Schr\"odinger states}\label{swf}

We first consider the orbital dynamics of a charged particle in a homogeneous magnetic field within the scalar Schr\"odinger theory, which provides the reference framework for the subsequent Pauli analysis.

For a homogeneous magnetic field $\bm B=B_0\bm e_z$, the Schr\"odinger equation reads
\begin{equation}\label{schrok}
i\hbar\partial_t\psi=\left[-\frac{\hbar^2}{2m}\left(\bm{\nabla}-\frac{iq}{\hbar}\bm A
\right)^2+cqA_0\right]\psi,
\end{equation}
where the gauge can be chosen such that
\begin{equation}\label{ceca}
A_0=0,\qquad \bm{A}=\left[0,B_0 x,0\right].
\end{equation}

Because vector potential (\ref{ceca}) is translationally invariant in $y$ and $z$ directions, it is natural to seek stationary solutions in the separated form
\begin{equation}\label{pozpi}
\psi(\bm{r},t)=e^{-i\omega t}e^{i(k_y y+k_z z)}\phi_0(x).
\end{equation}
The resulting eigenvalue problem reduces to that of a shifted one-dimensional harmonic oscillator. Introducing the quantities
\begin{equation}\label{pomx}
x_0=\frac{\hbar k_y}{q B_0},\quad \tilde{x}=x-x_0,\quad \Omega=\frac{qB_0}{m},
\end{equation}
where $\Omega$ is the signed cyclotron frequency, Eq.~(\ref{schrok}) takes the form
\begin{equation}\label{scha}
\left(-\frac{\hbar^2}{2m}\frac{\dd^2}{\dd\tilde{x}^2}+\frac{m\Omega^2\tilde{x}^2}{2}\right)\phi_0
=
\left(\hbar\omega-\frac{\hbar^2k_z^2}{2m}\right)\phi_0
\end{equation}
i.e., it is the standard harmonic-oscillator problem underlying Landau quantization \cite{ll,jl}. The transverse motion is quantized into discrete Landau levels, while motion along the magnetic-field direction remains free.

The lowest Landau level satisfies
\begin{equation}\label{engs}
\hbar\omega-\frac{\hbar^2k_z^2}{2m}=\frac{\hbar}{2}\,|\Omega|,
\end{equation}
and the corresponding wave-function reads
\begin{equation}\label{rozf}
\phi_0(x)=N\exp\left[-\frac{m|\Omega|}{2\hbar}(x-x_0)^2\right],
\end{equation}
with normalization
\begin{equation}\label{normn}
N=\left(\frac{m|\Omega|}{\pi \hbar}\right)^{1/4}.
\end{equation}

The parameter $x_0$ determines the guiding-center position associated with a given value of $k_y$. As a consequence, different transverse momenta correspond to identical Landau orbitals displaced along the $x$ direction. 

The corresponding probability current is
\begin{equation}\label{curschr}
\bm{j}=-\frac{i\hbar}{2m}\left(\psi^*\,\bm{\nabla}\psi-\bm{\nabla}\psi^*\, \psi\right)-\frac{q}{m}\,\psi^*\bm{A}\psi.
\end{equation}

Although the individual terms are gauge dependent, their sum is gauge invariant and therefore represents a physically observable probability flux.

\subsection{Pauli spinor states}\label{spwf}

We now extend the review to particles with intrinsic spin. In the nonrelativistic regime the dynamics is governed by the Pauli equation, which supplements the Schr\"odinger Hamiltonian with the coupling between spin and magnetic field. It may be obtained as the nonrelativistic limit of the Dirac theory \cite{fw,sakurai} in the form
\begin{equation}\label{pequ}
i\hbar\partial_t\Psi=\left[-\frac{\hbar^2}{2m}\left(\bm\nabla-\frac{iq}{\hbar}\bm A
\right)^2+cqA_0-\mu\bm\sigma\cdot\bm B\right]\Psi,
\end{equation}
where $\Psi$ is a two-component spinor.

For the uniform magnetic field (\ref{ceca}), stationary spinor solutions can be sought in the form
\begin{equation}\label{propf}
\Psi_s(\bm{r},t)=e^{-i\omega_s t}e^{i(k_y y+k_z z)}\phi_0(x)v_s,\quad s=\pm 1,
\end{equation}
with $\sigma_z v_s=s v_s$. Substitution into (\ref{pequ}) yields an equation with the same orbital structure as (\ref{scha}), the only modification being the Zeeman contribution:
\begin{equation}\label{roos}
\left(-\frac{\hbar^2}{2m}\frac{\dd^2}{\dd\tilde{x}^2}+\frac{m\Omega^2\tilde{x}^2}{2}\right)\phi_0
=
\Big(\hbar\omega_s-\frac{\hbar^2k_z^2}{2m}+s\mu B_0\Big)\phi_0.
\end{equation}

The orbital wave-function therefore coincides with that of the scalar case, while the energy relation becomes
\begin{equation}\label{engsp}
\hbar\omega_s-\frac{\hbar^2k_z^2}{2m}+s\mu B_0=\frac{\hbar}{2}\,|\Omega|.
\end{equation}

For an electron the magnetic moment equals $\mu=\frac{q\hbar}{2m}$. Equation
(\ref{engsp}) then yields
\begin{equation}
\omega_+-\omega_-=-\frac{2\mu B_0}{\hbar}=-\Omega.
\end{equation}
Thus, the relative spin phase evolves with the cyclotron frequency magnitude $|\Omega|$.

The Pauli probability current, which can be derived from the Dirac equation via the Gordon decomposition \cite{gordon,beli,takabayasi,shi}, in the nonrelativistic limit takes the form
\begin{equation}\label{pacur}
\bm j =-\frac{i\hbar}{2m}\left[\Psi^\dagger\bm\nabla\Psi
-\left(\bm\nabla\Psi^\dagger\right)\Psi\right]-\frac{q}{m}\bm A
\Psi^\dagger\Psi+\frac{\hbar}{m}\,\bm\nabla\times\bm S,
\end{equation}
where the dimensionless-spin density is defined as
\begin{equation}
\bm S =\Psi^\dagger\frac{\bm{\sigma}}{2}\Psi.
\end{equation}

Equation (\ref{pacur}) shows that the probability current consists of an orbital contribution identical to that of the Schr\"odinger theory and an additional spin current generated by the spin density. Consequently, interference effects may arise not only from orbital superpositions but also from the relative spin dynamics. As will be shown below, this additional structure provides a mechanism through which the magnetic field can influence and enhance probability backflow.

\section{Single-orbital current structure and constraints on backflow}\label{smb}

In this section we consider a single LLL state of the form (\ref{propf}), characterized by positive longitudinal momentum $k_z>0$ and a fixed guiding-center parameter $k_y$.

In the chosen gauge, the vector potential has no $z$ component, and therefore in the scalar case (\ref{pozpi}) the probability current along the magnetic-field direction is determined entirely by the longitudinal phase:
\begin{equation}\label{jza}
j_z=-\frac{i\hbar}{2m}\left(\psi^*\partial_z\psi-\partial_z\psi^*\psi\right)=\frac{\hbar k_z}{m}\,\psi^*\psi=\frac{\hbar k_z}{m}\,\phi_0^2.
\end{equation}

Since $k_z>0$ by definition, the longitudinal current is positive. As a result, a single Landau orbital does not exhibit probability backflow in the Schr\"odinger description. The magnetic field merely reshapes this probability distribution in the transverse plane via the guiding-center shift, but it does not introduce any mechanism capable of reversing the sign of the longitudinal current. Therefore, in this section we focus on the Pauli case, where the coupling between spin and magnetic field introduces the additional structure in the probability current and potentially opens the possibility of nontrivial competition between orbital and spin current contributions.

In the spinor case the $z$-component of the current reads
\begin{equation}\label{jzao}
j_z=-\frac{i\hbar}{2m}\left(\Psi^\dagger\partial_z\Psi-
\partial_z\Psi^\dagger\Psi\right)+ \frac{\hbar}{m}\left(\partial_xS_y-\partial_yS_x\right),
\end{equation}
i.e., it is a sum of the orbital ($j_z^{(o)}$) and spin ($j_z^{(s)}$) terms.

Let the initial spinor in (\ref{propf}) be chosen as a general normalized spin state
\begin{equation}\label{inispin}
v= \begin{pmatrix} a\\ b \end{pmatrix}, \qquad |a|^2+|b|^2=1.
\end{equation}
Because the Zeeman term splits spin-up and spin-down energies, as given in (\ref{engsp}), the two components evolve with different frequencies $\omega_\pm$ according to (\ref{propf}). Therefore, the exact one-mode Pauli spinor is
\begin{equation}\label{onemod}
\Psi(\bm{r},t)=e^{i(k_yy+k_zz)}\phi_0(x)
\begin{pmatrix} a\,e^{-i\omega_+ t}\\ b\,e^{-i\omega_- t} \end{pmatrix}.
\end{equation}

After substitution of (\ref{onemod}) into (\ref{jzao}) the orbital part becomes again
\begin{equation}\label{orbze}
j_z^{(o)}=\frac{\hbar k_z}{m}\,\phi_0^2,
\end{equation}
since $\partial_z\Psi=ik_z\Psi$. This component of the current, as in the scalar case, is strictly positive and is $B_0$-dependent only through the position of the Gaussian maximum.

Now let us turn to the spin component and first find the spin density:
\begin{eqnarray}\label{spinsd}
\bm S &=& \Psi^\dagger\frac{\bm\sigma}{2}\Psi=\frac{1}{2}\,\phi_0^2(x)\big[|a|^2v_+^\dagger\bm{\sigma}v_++|b|^2v_-^\dagger\bm{\sigma}v_-\nonumber\\
&&+a^*b\,e^{-i\Omega t}v_+^\dagger\bm{\sigma}v_-+ab^*\,e^{i\Omega t}v_-^\dagger\bm{\sigma}v_+\big].
\end{eqnarray}
It is straightforward to obtain that
\begin{subequations}\label{spicom}
\begin{align}
&v_+^\dagger\bm{\sigma}v_+=[0,0,1],\qquad v_-^\dagger\bm{\sigma}v_-=[0,0,-1],\label{spicom1}\\
&v_-^\dagger\bm{\sigma}v_+=[1,i,0],\qquad v_+^\dagger\bm{\sigma}v_-=[1,-i,0],\label{spicom2}
\end{align}
\end{subequations}
and after substitution into (\ref{spinsd}) the spin-density components are
\begin{subequations}\label{spcomp}
\begin{align}
S_x&=\frac{1}{2}\,\phi_0^2(x)\left(a^*b\,e^{-i\Omega t}+ab^*\,e^{i\Omega t}\right)\nonumber\\
&= \phi_0^2(x)\,\Re\!\left(a^*b\,e^{-i\Omega t}\right),\label{spcompx}\\
S_y&=\frac{1}{2i}\,\phi_0^2(x)\left(a^*b\,e^{-i\Omega t}-ab^*\,e^{i\Omega t}\right)\nonumber\\
&=\phi_0^2(x)\,\Im\!\left(a^*b\,e^{-i\Omega t}\right),\label{spcompy}\\
S_z&=\frac{1}{2}\,\phi_0^2(x)(|a|^2-|b|^2).\label{spcompz}
\end{align}
\end{subequations}
Thus, while the spin precesses in time, the spatial profile of the spin density remains unchanged and is governed solely by the factor $\phi_0^2(x)$. Thanks to this precession the magnetic field can potentially influence the interplay between orbital and spin currents, and hence the eventual backflow.

According to (\ref{jzao}) the spin contribution to $j_z$ reads
\begin{equation}
j_z^{(s)}=\frac{\hbar}{m}\left(\partial_xS_y-\partial_yS_x\right).
\end{equation}
Since the dependence on $y$ entirely disappeared, one has $\partial_yS_x=0$, and 
hence
\begin{equation}
j_z^{(s)}=\frac{\hbar}{m}\,\partial_xS_y=\frac{\hbar}{m}\,\partial_x\phi_0^2(x)\,\Im(a^*b\,e^{-i\Omega t}).
\end{equation}
Using (\ref{rozf}) one obtains
\begin{equation}
\partial_x\phi_0^2=-2\frac{m|\Omega|}{\hbar}(x-x_0)\phi_0^2.
\end{equation}
so the longitudinal spin current takes the form
\begin{equation}\label{djzs}
j_z^{(s)}=-2|\Omega|(x-x_0)\phi_0^2(x)\,\Im(a^*b\,e^{-i\Omega t}).
\end{equation}

The total current has therefore two components, of which the orbital contribution is strictly positive and the spin contribution is oscillating with time:
\begin{equation}\label{toca}
j_z=\phi_0^2(x)\left[\frac{\hbar k_z}{m}-2|\Omega|(x-x_0)\Im(a^*b\,e^{-i\Omega t})\right].
\end{equation}
Now writing $a^*b=|ab|e^{i\varphi}$, one has
\begin{equation}
\Im(a^*b\,e^{-i\Omega t})=-|ab|\sin(\Omega t-\varphi),
\end{equation}
and then
\begin{equation}\label{tocar}
j_z=\phi_0^2(x)\left[\frac{\hbar k_z}{m}+2|\Omega|(x-x_0)\sin(\Omega t-\varphi)\right].
\end{equation}

This form shows that the longitudinal current acquires a time-dependent contribution whose amplitude is controlled by the magnetic field through $|\Omega|$, while its temporal structure is governed by the relative spinor phase encoded in $\varphi$. The spatial dependence enters via the Landau guiding-center shift $(x-x_0)$, which determines the strength of the spin contribution through the local weight of the Gaussian envelope.

The structure of Eq.~(\ref{tocar}) makes it clear that the effect does not originate from interference between different momentum components in the usual sense. Instead, it arises from the spin-dependent contribution to the Pauli probability current in Eq.~(\ref{pacur}), which couples longitudinal transport to the internal spin dynamics in the presence of a magnetic field. Unlike conventional quantum backflow generated in momentum space, here the relevant phase structure is carried by the spinor degrees of freedom and their relative evolution, while the orbital motion remains confined within a single Landau mode.

This suggests that the magnetic field could provide a direct control parameter for the backflow, since oscillation frequency, spin-current amplitude and shifted coordinate depend explicitly on $B_0$. A closer inspection shows, however, that this control is strongly constrained.

A functional measure of the amount of backflow at fixed position $x$ may be defined \cite{yearsley,trpauli} as
\begin{equation}\label{bfunc}
F(T)=\int\limits_{t_0-\frac{T}{2}}^{t_0+\frac{T}{2}}\hspace{-1ex}\dd t\,j_z(t),
\end{equation}
where $t_0$ is chosen to maximize the effect. It should be noted that, unlike the standard one-dimensional backflow problem, the present functional is inherently local because the probability current is spatially confined by the Landau orbitals. The local time-integrated current is therefore used as a convenient quantity for analyzing magnetic-field-controlled backflow. For a physically realistic detector, however, the current must also be integrated over its finite transverse cross section, in addition to the time interval. Such spatial averaging is not merely a mathematical smoothing, but an essential part of the measurement process. The resulting space-time integrated current provides a physically accessible, dimensionless measure of the probability transfer, and its robustness is examined in Sec. \ref{res}.

\begin{figure*}[t]
\centering
\includegraphics[width=0.69\textwidth,angle=0]{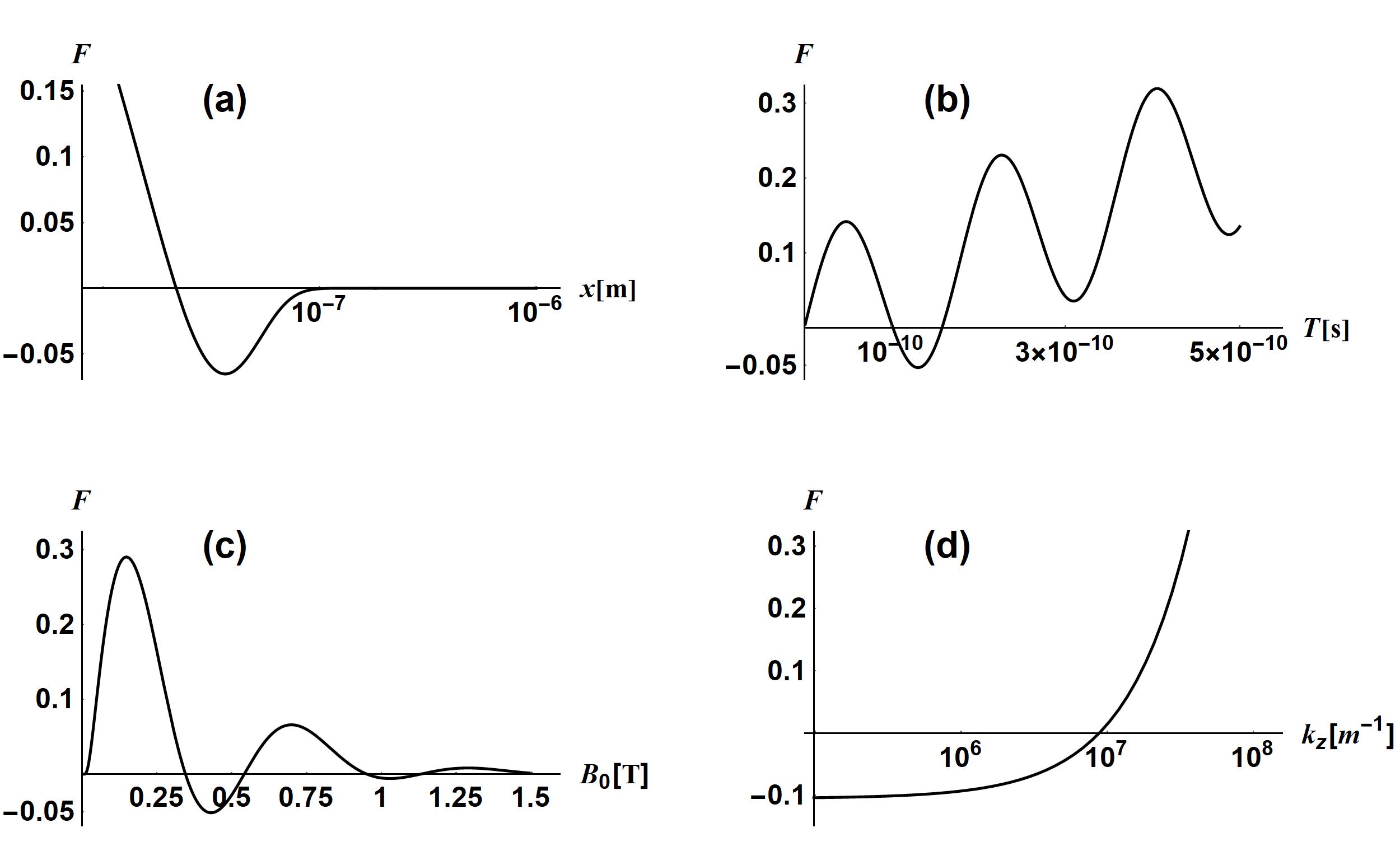}
\caption{Dependence of the backflow functional $F$ in the single-orbital regime.
Unless stated otherwise, the parameters are fixed to
$x=5\times 10^{-8}\,\mathrm{m}$, $T=1.2\times 10^{-10}\,\mathrm{s}$, $B_0=0.4\,\mathrm{T}$,
and $k_z=0.5\times 10^{7}\,\mathrm{m}^{-1}$.
(a) $F$ as a function of the spatial coordinate $x$, with all other parameters fixed.
(b) $F$ as a function of the observation time $T$ at fixed $x$, $B_0$, and $k_z$.
(c) $F$ as a function of the magnetic field strength $B_0$ at fixed $x$, $T$, and $k_z$.
(d) $F$ as a function of longitudinal momentum $k_z$ at fixed $x$, $T$, and $B_0$.}
\label{onemode}
\end{figure*}

Evaluating the time integral and using the identity for the difference of two cosines, one obtains for $F(T)$
\begin{eqnarray}\label{fta}
F(T)&=&\phi_0^2(x)\bigg[\frac{\hbar k_z}{m}\,T\\
&+&4\,\frac{|\Omega|}{\Omega}\,|ab|(x-x_0)\sin(\Omega t_0-\varphi)\sin\tfrac{\Omega T}{2}\bigg].\nonumber
\end{eqnarray}
The parameter $t_0$ is treated as a freely adjustable reference time and determines the center of the optimal measurement window. It is chosen such that the oscillatory contribution is maximized, i.e.\ $\sin(\Omega t_0-\varphi)=1$:
\begin{equation}\label{nkw}
t_0=\frac{\pi/2+\varphi}{\Omega}.
\end{equation}
The spinor coefficient $|ab|$ with the normalization $|a|^2+|b|^2=1$, in turn, is maximized for $|a|=|b|=\frac{1}{\sqrt{2}}$. In this way the final expression is obtained in the form
\begin{equation}\label{singlef}
F(T)\!=\!\phi_0^2(x)\,\frac{\hbar k_z}{m}\,T\left[1+\frac{m|\Omega|(x-x_0)}{\hbar k_z}\cdot
\frac{\sin\tfrac{\Omega T}{2}}{\tfrac{\Omega T}{2}}\right]\!.
\end{equation}
Equation (\ref{singlef}) shows that magnetic-field control enters through a spin-precession correction multiplied by the Gaussian envelope of the same orbital mode.

The condition for the dominance of the spin contribution, at least for specifically chosen time windows, reads
\begin{equation}
\alpha:= \frac{m|\Omega|\,|x-x_0|}{\hbar k_z} > 1.
\end{equation}
In the regime of short observation times $T \ll |\Omega|^{-1}$, the oscillatory factor can be approximated as
\begin{equation}\label{ttr}
\frac{\sin\tfrac{\Omega T}{2}}{\tfrac{\Omega T}{2}}\approx 1,
\end{equation}
so that temporal averaging does not significantly suppress the effect. Within this regime the quantity $\alpha$ controls the relative magnitude of the spin-induced term.
However, (\ref{ttr}) cannot be achieved by reducing the magnetic field $B_0$ (and thus increasing $|\Omega|^{-1}$), since in this limit the amplitude of the spin-induced contribution is simultaneously suppressed and no backflow is generated within a single-mode configuration.

The parameter $\alpha$ controls not only the relative magnitude of the spin contribution with respect to the orbital current, but also fixes the spatial point at which this contribution is effectively probed. In particular, the condition $\alpha>1$, required for the spin term to dominate, necessarily shifts the evaluation point away from the center of the wavepacket into its exponentially suppressed tail. This follows from the relation
\begin{equation}\label{shiftx}
|x-x_0|=\alpha\,\frac{\hbar k_z}{m|\Omega|}.
\end{equation}

At this shifted position, the Gaussian profile is strongly reduced. The suppression factor is obtained from the ratio of the envelope at $x$ to its maximal value,
\begin{equation}\label{stob}
\left(\frac{\phi_0(x)}{\phi_{0\,\text{max}}}\right)^2=
\exp\!\left[-\frac{\hbar k_z^2}{m|\Omega|}\,\alpha^2\right]=
\exp\!\left[-\frac{\hbar k_z^2}{|qB_0|}\,\alpha^2\right].
\end{equation}

Thus, increasing $\alpha$ produces a competing effect: while it enhances the spin-to-orbital ratio, it simultaneously reduces the local probability density through exponential damping of the envelope.

For example for electrons, when $B_0=1\,\mathrm{T}$ and $k_z=10^8\,\mathrm{m^{-1}}$, a representative value $\alpha \approx 1.5$ yields
\begin{equation}\label{stoba}
\left(\frac{\phi_0(x)}{\phi_{0\,\text{max}}}\right)^2
\approx (0.0019)^{\alpha^2}
\approx 7.8\cdot 10^{-7}.
\end{equation}
so in this regime the signal is strongly suppressed. The net backflow effect---although existent---is effectively attenuated under physically realistic conditions. The value of (\ref{stoba}) can however be strongly increased by considering very cold electrons (at least as regards their longitudinal motion). For instance if $k_z=10^7\,\mathrm{m^{-1}}$ (which corresponds to $3.4\,\mu\mathrm{eV}$ of kinetic energy) the factor (\ref{stoba}) equals $0.87$ making the effect observable and proving that the presence of a magnetic field allows the spin alone to generate negative probability flow even in one-mode configuration.

One should mention that the parameter $k_z$ plays a dual role in the above mechanism.
Decreasing $k_z$ reduces the spatial displacement required for the spin contribution to dominate, thereby weakening the Gaussian suppression. On the other hand, excessively small values of $k_z$ correspond to extremely low kinetic energies (connected with the longitudinal motion) and become difficult to justify experimentally. The value $k_z\sim 10^7\,\mathrm{m^{-1}}$ can represent a reasonable compromise between these competing effects. One can conclude that in a single-orbital setting the constant magnetic field aligned along the propagation axis provides some kind of practical control of the backflow but limited to the case of slow longitudinal motion.

Figure~\ref{onemode} summarizes the parameter dependence of the backflow functional $F(T)$ in the single-orbital case. Panel (a) shows $F$ as a function of the observation point $x$ for fixed $T$, $B_0$ and $k_z$. The backflow exhibits a pronounced spatial localization: it reaches its maximal magnitude at a well-defined position $x \simeq 5\times 10^{-8}\,\mathrm{m}$ for $B_0=0.4\,\mathrm{T}$, while decaying rapidly away from this point partially due to the Gaussian-Landau envelope.

Panel (b) displays the dependence on the observation time $T$, revealing the existence of an optimal temporal window. For the same magnetic field strength and optimal $x$, the backflow is maximized around $T \approx 1.2\times 10^{-10}\,\mathrm{s}$, with suppression occurring for both shorter and longer integration times due to the oscillatory spin-precession factor.

The magnetic-field dependence shown in panel (c) indicates a pronounced optimal-field behaviour: the backflow attains its maximum at an intermediate field strength $B_0 \approx 0.4\,\mathrm{T}$, while weaker and stronger fields reduce the effect due to competing $B_0$-dependent mechanisms. On the one hand, the spin-induced contribution grows with $|\Omega| \propto B_0$, whereas increasing the field strength modifies the spatial localization of the Landau state and shifts the relevant sampling region into a regime where the Gaussian envelope suppresses the signal.

Finally, panel (d) demonstrates a sharp threshold behaviour with respect to the longitudinal momentum. For $k_z \lesssim 10^{7}\,\mathrm{m^{-1}}$ the backflow remains appreciable, whereas above this scale the effect is strongly suppressed and effectively disappears.

Overall, the figure reveals that single-orbital backflow is not a generic feature, but rather a finely tuned effect requiring simultaneous optimization in space, time, magnetic field strength, and longitudinal kinetic energy.

\section{Two-mode quantum interference and optimal magnetic control of backflow}\label{tmqi}

As we have seen, the spin-induced backflow in a single Landau orbital remains limited even for optimally chosen magnetic fields. In contrast, pronounced quantum backflow is typically associated with interference between distinct modes. We therefore consider the minimal configuration that can exhibit such effects, namely a superposition of two Landau states, and investigate it in both the Schr\"odinger and Pauli theories. This framework allows one to separate orbital and spin contributions to the current and to determine how the magnetic field influences the backflow phenomenon.

\subsection{Two-mode backflow in the Schr\"odinger theory}

As seen in Sec.~\ref{smb}, a single scalar Landau orbital cannot exhibit backflow. The minimal configuration capable of generating negative probability flux therefore involves quantum interference between at least two distinct modes \cite{brack,yearsley,muga}. Hence, a superposition of two LLL states is considered below:
\begin{equation}\label{podp}
\psi=c_1\psi_1+c_2\psi_2,\qquad |c_1|^2+|c_2|^2=1.
\end{equation}

Each component is taken in the form
\begin{equation}\label{pozpip}
\psi_i(\bm{r},t)=e^{-i\omega^{(i)} t}e^{i(k^{(i)}_y y+k^{(i)}_z z)}
\phi_{0i}(x),\qquad i=1,2,
\end{equation}
with the corresponding $i$-th LLL orbital
\begin{equation}\label{rozfy}
\phi_{0i}(x)=N\exp\left[-\frac{m|\Omega|}{2\hbar}(x-x_{0i})^2\right],
\end{equation}
where $N$ is given by (\ref{normn}) and
\begin{equation}\label{pomxi}
x_{0i}=\frac{\hbar k^{(i)}_y}{q B_0}.
\end{equation}

It is convenient to introduce the guiding-center separation
and the midpoint coordinate,
\begin{subequations}\label{rox}
\begin{align}
\Delta x&=x_{02}-x_{01}=\frac{\hbar \left(k^{(2)}_y-k^{(1)}_y\right)}{q B_0},
\label{rox1}\\
x_c&=\frac{x_{01}+x_{02}}{2}=\frac{\hbar \left(k^{(1)}_y+k^{(2)}_y\right)}{2q B_0}.
\label{rox2}
\end{align}
\end{subequations}

\begin{figure}[h!]
\centering
\includegraphics[width=0.38\textwidth,angle=0]{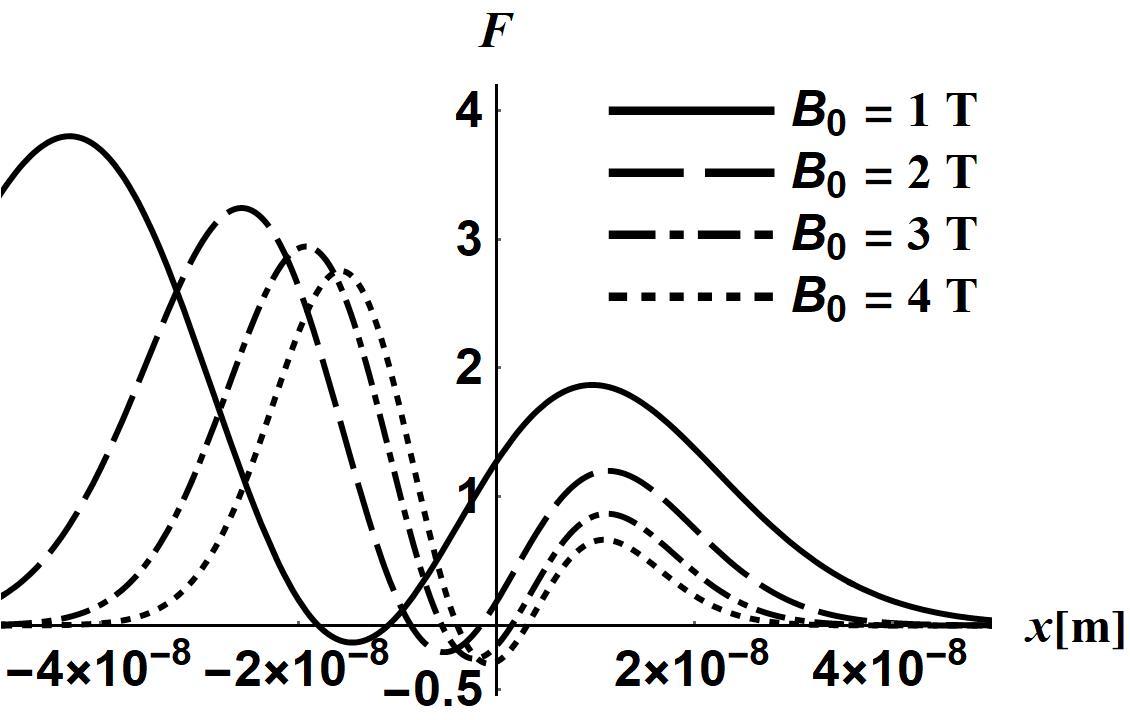}
\caption{Dependence of the backflow functional on the magnetic field strength. The four curves correspond to $B_0 = 1,2,3,4\,\mathrm{T}$ for fixed observation time $T=6\times 10^{-9}\,\mathrm{s}$ and momenta: $k_z^{(1)}=10^6\,\mathrm{m}^{-1}$, $k_z^{(2)}=1.4\times 10^6\,\mathrm{m}^{-1}$, $k_y^{(1)}=10^7\,\mathrm{m}^{-1}$ and $k_y^{(2)}=5\times 10^7\,\mathrm{m}^{-1}$. As the magnetic field increases, the region of negative probability flux shifts towards $x=0$, reflecting the $1/|qB_0|$ scaling of the guiding-center locations and the concomitant reduction of the spatial displacement between the interfering Landau orbitals.}
\label{twomode}
\end{figure}

The quantity $x_c$ defines the center of the spatial overlap region between Landau orbitals. For electrons ($q=-e$) and $B_0>0$, one has $x_c<0$ whenever $k_y^{(1)}+k_y^{(2)}>0$. Consequently, the region where the orbital overlap—and hence the interference contribution—is expected to be the strongest is naturally shifted towards negative values of $x$. This behavior is illustrated later in Fig.~\ref{twomode}.

The two modes may differ both in their longitudinal momenta $k_z^{(i)}$ and in the guiding-center positions $x_{0i}$ determined by the transverse momenta $k_y^{(i)}$. The difference $\Delta x$ determines the spatial overlap of the Landau orbitals and hence the magnitude of the interference contribution. Substituting (\ref{podp}) into the Schr\"odinger current (\ref{curschr}) and taking the longitudinal component yields
\begin{eqnarray}\label{jzas}
j_z&=&-\frac{i\hbar}{2m}\Big[\left(c_1^*\psi_1^*+c_2^*\psi_2^*\right)
\partial_z\left(c_1\psi_1+c_2\psi_2\right)\nonumber\\
&&-\partial_z\left(c_1^*\psi_1^*+c_2^*\psi_2^*\right)
\left(c_1\psi_1+c_2\psi_2\right)\Big].
\end{eqnarray}

To maximize the interference contribution for a fixed normalization, equal amplitudes with a relative phase of $\pi$ are chosen:
\begin{equation}\label{secc}
c_1=\frac{1}{\sqrt2}=-c_2.
\end{equation}
Under the normalization condition (\ref{podp}), equal amplitudes maximize $|c_1c_2|$. The phase can be fixed as in (\ref{secc}) without loss of generality, as any constant phase shift can be absorbed into the interference phase. The probability current then takes the form
\begin{eqnarray}\label{prajz}
j_z&=&\frac{\hbar}{2m}\Big[k_z^{(1)}\phi_{01}^2+k_z^{(2)}\phi_{02}^2-\phi_{01}\phi_{02}\left(k_z^{(1)}+k_z^{(2)}\right)\nonumber\\
&&\times \cos(\Delta\omega t-\Delta k_y y-\Delta k_z z)\Big].
\end{eqnarray}
The differences in frequency and wave vectors entering the interference phase are defined as
\begin{subequations}\label{rodel}
\begin{align}
&\Delta\omega =\omega^{(2)}-\omega^{(1)}=\frac{\hbar}{2m}\left[(k_z^{(2)})^2-(k_z^{(1)})^2\right],\label{rodela}\\
&\Delta k_y=k_y^{(2)}-k_y^{(1)},\qquad \Delta k_z=k_z^{(2)}-k_z^{(1)}.\label{rodelb}
\end{align}
\end{subequations}

It is important to note that $\Delta\omega$ originates purely from the longitudinal kinetic dispersion. Since both modes belong to the same Landau level, they possess identical energies associated with the motion in the plane perpendicular to the magnetic field. They therefore cancel in $\omega^{(2)}-\omega^{(1)}$, leaving only the longitudinal kinetic term.

The interference pattern is determined by the phase differences between the two modes: $\Delta k_y$ and $\Delta k_z$ set the spatial modulation transverse and parallel to the propagation axis, respectively, while $\Delta\omega$ determines its temporal variation. Negative current may arise when the interference term locally dominates the positive contributions of the individual modes. Unlike in free space, Landau localization shapes the interference pattern in the plane perpendicular to the magnetic field, making the effective overlap region field dependent.

The backflow functional is evaluated according to (\ref{bfunc}). Performing the time integration over an interval centered at certain $t_0$ produces the standard $\sin(\Delta\omega T/2)/(\Delta\omega/2)$ factor together with a phase-dependent cosine term evaluated at the center of the integration window:
\begin{eqnarray}\label{fsdwa}
F(T)&=&\frac{\hbar}{2m}\Big[\left(k_z^{(1)}\phi_{01}^2+k_z^{(2)}\phi_{02}^2\right)T-2\phi_{01}\phi_{02}\,\frac{k_z^{(1)}+k_z^{(2)}}{\Delta \omega}\nonumber\\
&&\times \cos(\Delta\omega t_0-\Delta k_y y-\Delta k_z z)\,
\sin\frac{\Delta\omega T}{2}\Big].
\end{eqnarray}

For any fixed values of $y$ and $z$, the optimal observation time $t_0$ can be chosen such that the cosine factor is maximized. The finite instrumental resolution required to realize this optimal alignment will be discussed later.

Thus the backflow functional becomes
\begin{eqnarray}\label{fsdwas}
F(T)&=&\frac{\hbar}{2m}\Big[\left(k_z^{(1)}\phi_{01}^2+k_z^{(2)}\phi_{02}^2\right)T\\
&&-\phi_{01}\phi_{02}\,\frac{4m }{\hbar\Delta k_z} \sin\frac{\Delta\omega T}{2}\Big].\nonumber
\end{eqnarray}
The interference term depends solely on the longitudinal momentum detuning $\Delta k_z$, while the average longitudinal momentum cancels out upon using the dispersion relation. We assume $k_z^{(1)} \neq k_z^{(2)}$ and, without loss of generality, order the two modes such that $k_z^{(1)} < k_z^{(2)}$.

Let us now analyze whether the interference contribution can dominate over the positive orbital part for certain values of $T$. Negative values of $F(T)$ arise in regions where, for a fixed $x$, the following condition is satisfied:
\begin{equation}\label{nefte}
k_z^{(1)}\phi_{01}(x)^2+k_z^{(2)}\phi_{02}(x)^2<\frac{4m}{\hbar\Delta k_z T}\,\phi_{01}(x)\phi_{02}(x).
\end{equation}

The point $x=x_c$ does not, in general, maximize the backflow signal. At $x=x_c$ one has $\phi_{01}(x_c)=\phi_{02}(x_c)$, which maximizes the overlap factor $\phi_{01}\phi_{02}$, but the left-hand side remains asymmetrically weighted by $k_z^{(1)}, k_z^{(2)}$. 

For a given position $x$, let us introduce the ratio
\begin{equation}\label{defr}
r(x)=\frac{\phi_{01}(x)}{\phi_{02}(x)}=
e^{-\frac{m|\Omega|}{\hbar}(x-x_c)\Delta x}=e^{-\operatorname{sgn}(qB_0)\,\Delta k_y (x-x_c)},
\end{equation}
which is strictly positive and varies monotonically with $x$, covering the full range $r(x)\in(0,\infty)$. 
In terms of $r$, condition (\ref{nefte}) becomes the quadratic inequality
\begin{equation}\label{rokr}
k_z^{(1)} r^2-\frac{4m}{\hbar\Delta k_z T}\,r+k_z^{(2)}<0.
\end{equation}

The existence of backflow is determined by whether this quadratic polynomial becomes negative for some positive $r$. Since $r(x)$ spans $(0,\infty)$, any admissible root structure can be realized at an appropriate position $x$. The discriminant of (\ref{rokr}) is
\begin{equation}\label{discr}
\Delta= \frac{16 m^2}{\hbar^2\Delta k_z^2 T^2}-4k_z^{(1)}k_z^{(2)}.
\end{equation}

If $\Delta>0$, the quadratic polynomial has two real roots, both positive because all coefficients in (\ref{rokr}) are positive except the linear term. Hence, the inequality (\ref{rokr}) is satisfied over an interval of $r>0$, corresponding to a spatial region where backflow occurs. Setting $k_z^{(2)}=a k_z^{(1)}$ with $a>0$, the condition $\Delta>0$ can be rewritten as
\begin{equation}\label{disca}
a\, (k_z^{(1)})^2\,T^2 < \frac{4m^2}{\hbar^2\Delta k_z^2},\;\;\;\;\text{or}\;\;\;\;
a (a-1)^2< \frac{4m^2}{\hbar^2 (k_z^{(1)})^4 T^2}.
\end{equation}
The second condition is fulfilled for a range of physically accessible parameters. The left-hand side decreases with decreasing longitudinal momentum detuning $\Delta k_z$, while the right-hand side increases as the observation time $T$ is reduced, within the validity range of the single-frequency approximation and finite measurement resolution.

The condition for the existence of backflow is independent of the magnetic field strength $B_0$. This reflects the fact that, within the LLL approximation, the magnetic field affects only the transverse localization of the wavefunctions, while the interference responsible for backflow is governed entirely by longitudinal momentum detuning.

Although the existence criterion for backflow is independent of $B_0$, its spatial localization and magnitude are not. The field enters through the guiding-center positions (\ref{pomxi}), which depend on the combination $\mathcal{B}=qB_0$. Consequently, the characteristic separation of the interference region scales as $1/|\mathcal{B}|$, so that in the weak-field regime the region supporting backflow is displaced to increasingly large distances from the origin.

At the same time, the Gaussian width of the LLL orbitals scales as $|\mathcal{B}|^{-1/2}$, corresponding to the magnetic length $\ell_B=\sqrt{\hbar/|qB_0|}$. Hence the spatial extent of the wavepackets grows more slowly than the separation of their centers. In the limit $|\mathcal{B}|\to 0$ the interference region becomes effectively delocalized and practically inaccessible, despite the formal persistence of the backflow condition. This shows that from the practical point of view the magnetic field is crucial for the appearance of backflow.
The optimal value of $B_0$ follows from maximizing the effective envelope function
\begin{equation}\label{fbb}
f(|\mathcal{B}|)\sim |\mathcal{B}|^{1/2}
\exp\!\left[-A|\mathcal{B}|-\frac{C}{|\mathcal{B}|}\right],
\end{equation}
or, equivalently, by maximizing its logarithm, where
\begin{equation}
A=\frac{x^2}{\hbar},\qquad
C=\frac{\hbar}{2}\big[(k_y^{(1)})^2+(k_y^{(2)})^2\big].
\end{equation}

The extremum condition yields
\begin{equation}\label{adb}
A|\mathcal{B}|^2-\frac{1}{2}|\mathcal{B}|-C=0,
\end{equation}
with the solution
\begin{eqnarray}\label{aop}
|\mathcal{B}|_{\mathrm{opt}}&=&\frac{\frac{1}{2}+\sqrt{\frac{1}{4}+4AC}}{2A}\\
&=&\frac{\hbar}{x^2}\left[1+\sqrt{1+8x^2\left((k_y^{(1)})^2+(k_y^{(2)})^2\right)}\right],\nonumber
\end{eqnarray}
and $|B_0|^{\mathrm{opt}}=|\mathcal{B}_{\mathrm{opt}}/q|$.

Thus, the optimal field magnitude is obtained by maximizing the effective overlap function (\ref{fbb}), which captures the competition between magnetic confinement, enhancing localization with increasing $|\mathcal{B}|$, and spatial separation of the guiding centers, suppressing overlap in the weak-field regime. The prefactor $|\mathcal{B}|^{1/2}$ reflects the scaling of the magnetic length and ensures a finite optimum. Since the overlap depends on $x$, the optimal field obtained from local maximization acquires a parametric $x$-dependence, indicating the fact that different spatial regions contribute most strongly to the interference signal at different field strengths.

Figure~\ref{twomode} shows the backflow functional for $B_0 = 1,2,3,4,\mathrm{T}$ at fixed observation time and longitudinal momenta. With increasing field, the region of negative probability flux shifts towards the origin, reflecting the $1/|qB_0|$ scaling of the guiding-center locations and the $|qB_0|^{-1/2}$ scaling of the Landau width. Despite this compression, the interference amplitude remains finite due to the overlap of the two Landau orbitals.

\subsection{Two-mode backflow in Pauli theory}

We now consider a genuine two-mode configuration involving both spin and spatial degrees of freedom. In this setting, the magnetic field $B_0$ plays a twofold role:
\begin{itemize}
\item it controls the spatial overlap of Landau orbitals via the guiding-center separation,
\item it determines the spin precession frequency and hence the temporal interference structure.
\end{itemize}
As a result, $B_0$ acts as a true control parameter of the probability current rather than a passive external field.

Let us consider two modes of the form
\begin{eqnarray}\label{2modes}
\Psi_i(\bm r,t)&=&e^{i(k^{(i)}_yy+k^{(i)}_zz)}\phi_{0i}(x)
\begin{pmatrix} a_i\,e^{-i\omega^{(i)}_+ t}\\ b_i\,e^{-i\omega^{(i)}_- t} \end{pmatrix}\\
&=:&e^{i(k_y^{(i)}y+k_z^{(i)}z)}\phi_{0i}(x)\chi_i(t),\quad i=1,2.\nonumber
\end{eqnarray}
where $|a_i|^2+|b_i|^2=1$, and in general
\begin{equation}
k_y^{(1)}\neq k_y^{(2)},\qquad k_z^{(1)}\neq k_z^{(2)}.
\end{equation}
The functions $\phi_{0i}(x)$ are defined in (\ref{rozfy}) and $\Delta k_y,\Delta k_z$ in (\ref{rodelb}). Since both modes experience the same Zeeman splitting, the spin-dependent
energy shift cancels when the frequency difference between the modes is
formed. Consequently, 
\begin{equation}
\Delta\omega_\pm =\omega^{(2)}_\pm-\omega^{(1)}_\pm
=\frac{\hbar}{2m}\left[(k_z^{(2)})^2-(k_z^{(1)})^2\right]=\Delta\omega.
\end{equation}
Thus, the intermode interference frequency is determined solely by the
difference of longitudinal kinetic energies. The Zeeman splitting remains
present and generates the spin-precession frequency $\Omega$, but it does
not contribute to $\Delta\omega$.

The full Pauli state is a superposition
\begin{equation}\label{twophi}
\Psi(\bm r,t)=c_1\Psi_1(\bm r,t)+c_2\Psi_2(\bm r,t),
\qquad |c_1|^2+|c_2|^2=1.
\end{equation}
The probability current splits into orbital and spin contributions. The former reads
\begin{eqnarray}\label{jzak}
j^{(o)}_z&=&-\frac{i\hbar}{2m}\Big[\Big(c_1^*\Psi_1^\dagger+c_2^*\Psi_2^\dagger\Big)\partial_z\Big(c_1\Psi_1+c_2\Psi_2\Big)\nonumber\\
&&-\partial_z\Big(c_1^*\Psi_1^\dagger+c_2^*\Psi_2^\dagger\Big)\Big(c_1\Psi_1+c_2\Psi_2\Big)\Big],
\end{eqnarray}
and contains both diagonal and interference terms as in the Schr\"odinger case. Using
\begin{equation}\label{popoz}
\partial_z\Psi_i(\bm r,t)=ik_z^{(i)}\Psi_i(\bm r,t),
\end{equation}
and
\begin{equation}\label{gest}
\Psi_1^\dagger(\bm r,t)\Psi_2(\bm r,t)=e^{i(\Delta k_y y+\Delta k_z z-\Delta\omega t)}
\phi_{01}(x)\phi_{02}(x)\Gamma_0,
\end{equation}
where $\Gamma_0=a_1^*a_2+b_1^*b_2$, we obtain
\begin{eqnarray}\label{orbtwo}
j_z^{(o)}&=&\frac{\hbar}{m}\Big[k_z^{(1)}|c_1|^2\phi_{01}^2+k_z^{(2)}|c_2|^2\phi_{02}^2+\left(k_z^{(1)}+k_z^{(2)}\right)\nonumber\\
&&\times\phi_{01}\phi_{02}\Re\left(c_1^*c_2 e^{i(\Delta k_y y+\Delta k_z z-\Delta\omega t)}\Gamma_0\right)\Big].
\end{eqnarray}

To determine the spin contribution, we next evaluate
the spin density entering the magnetization term of the Pauli current:
\begin{eqnarray}\label{scomp}
\bm{S}&\!=&\!\left(c_1^*\Psi_1^\dagger\! +\! c_2^*\Psi _2^\dagger\right)\frac{\bm{\sigma}}{2}\left(c_1\Psi _1\! +\! c_2\Psi _2\right)
= |c_1|^2 \phi_{01}^2\bm{\Gamma}_1(t)\\
&&\!\!\!\!\!\!\!\!\!+|c_2|^2 \phi_{02}^2\bm{\Gamma}_2(t)+\Re\left(c_1^*c_2 e^{i(\Delta k_y y+\Delta k_z z-\Delta\omega t)}\phi_{01}\phi_{02}\,\bm{\Sigma}\right).\nonumber
\end{eqnarray}
The vectors $\bm{\Gamma}_i$ and $\bm{\Sigma}$ are obtained by evaluating the
diagonal and off-diagonal matrix elements
$\chi_i^\dagger(\bm{\sigma}/2)\chi_i$ and
$\chi_1^\dagger(\bm{\sigma}/2)\chi_2$, respectively.
For $\bm{\Gamma}_i$ one obtains
\begin{equation}\label{gammy}
\bm{\Gamma}_i= \left[\Re(a_i^*b_ie^{-i\Omega t}),\Im(a_i^*b_ie^{-i\Omega t}),
\frac{1}{2}(|a_i|^2-|b_i|^2)\right],
\end{equation}
and the spin-interference vector is
\begin{eqnarray}\label{sissi}
\bm{\Sigma}&=&\Big[a_1^*b_2e^{-i\Omega t}+b_1^*a_2e^{i\Omega t},\\
&&-i a_1^*b_2e^{-i\Omega t}+i b_1^*a_2e^{i\Omega t},a_1^*a_2-b_1^*b_2\Big].\nonumber
\end{eqnarray}

Finally, the longitudinal component of the spin current decomposes into three terms which can be interpreted as local spin transport, gradient-induced overlap, and two-mode interference:
\begin{equation}\label{jsz}
j_z^{(s)}=\frac{\hbar}{m}\left(\partial_xS_y-\partial_yS_x\right)
=j_{z,\mathrm{loc}}^{(s)}+j_{z,\mathrm{grad}}^{(s)}+j_{z,\mathrm{int}}^{(s)},
\end{equation}
where
\begin{subequations}\label{jszc}
\begin{align}
&j_{z,\mathrm{loc}}^{(s)}=\frac{\hbar}{m}\sum_{i=1}^2|c_i|^2\left(\partial_x\phi_{0i}^2\right)\Gamma_{yi},\label{jszca}\\
\displaybreak
&j_{z,\mathrm{grad}}^{(s)}=\frac{\hbar}{m}\partial_x\left(\phi_{01}\phi_{02}\right)\Re\left[c_1^*c_2e^{i(\Delta k_y y+\Delta k_z z-\Delta\omega t)}
\Sigma_{y}\right],\label{jszcb}\\
&j_{z,\mathrm{int}}^{(s)}=\frac{\hbar\Delta k_y}{m}\,\phi_{01}\phi_{02}\Im\left[c_1^*c_2e^{i(\Delta k_y y+\Delta k_z z-\Delta\omega t)}\Sigma_{x}\right].\label{jszcc}
\end{align}
\end{subequations}
The term $j_{z,\mathrm{int}}^{(s)}$ arises only in two-mode configuration.
It originates from the transverse phase gradient associated with
$\Delta k_y$ and has no analogue in the single-orbital case.
This decomposition clearly separates three mechanisms contributing to
the spin current: local spin precession within each orbital, modulation of the
orbital overlap by spin dynamics, and genuine two-mode spin interference.

The interference terms are most pronounced for $|c_1|=|c_2|$, so we choose
\begin{equation}\label{cc12}
c_1=e^{-i\alpha} c_2=\frac{1}{\sqrt{2}}.
\end{equation}
To reduce the number of free parameters while retaining the essential interference mechanisms, we also fix the spinors as
\begin{equation}\label{wrca}
a_1=a_2=\frac{1}{\sqrt{2}},\qquad b_1=b_2=e^{i\chi}\,\frac{1}{\sqrt{2}}.
\end{equation}
The phases $\alpha$ and $\chi$ will be specified below. With this choice one finds
\begin{subequations}\label{gamust}
\begin{align}
&\Gamma_0=1,\label{gamust0}\\
&\bm{\Gamma} := \bm{\Gamma}_i=\frac{1}{2}\big[\cos(\Omega t-\chi),-\sin(\Omega t-\chi),0\big]=\frac{1}{2}\,\bm{\Sigma}. \label{gamust1}
\end{align}
\end{subequations}
This parametrization considerably simplifies the subsequent analysis without affecting the essential physical mechanisms responsible for backflow. The contributions to the longitudinal current now take the form
\begin{subequations}\label{jszupr}
\begin{align}
&j_z^{(o)}=\frac{\hbar}{2m}\Big[k_z^{(1)}\phi_{01}^2+k_z^{(2)}\phi_{02}^2\label{jszupro}\\
&\hspace{2ex}+\left(k_z^{(1)}+k_z^{(2)}\right)\phi_{01}\phi_{02}\cos(\Delta k_y y+\Delta k_z z-\Delta\omega t+\alpha)\Big].\nonumber\\
&j_{z,\mathrm{loc}}^{(s)}=-\frac{\hbar}{2m}\partial_x\left(\phi_{01}^2+\phi_{02}^2\right)\sin(\Omega t-\chi)\label{jszuprl}\\
&j_{z,\mathrm{grad}}^{(s)}=-\frac{\hbar}{2m}\partial_x\left(\phi_{01}\phi_{02}\right)\label{jszuprg}\\
&\hspace{9ex}\times\cos(\Delta k_y y+\Delta k_z z-\Delta\omega t+\alpha)\sin(\Omega t-\chi),\nonumber\\
&j_{z,\mathrm{int}}^{(s)}=\frac{\hbar\Delta k_y}{2m}\,\phi_{01}\phi_{02}\label{jszupri}\\
&\hspace{7ex}\times\sin(\Delta k_y y+\Delta k_z z-\Delta\omega t+\alpha)\cos(\Omega t-\chi).\nonumber
\end{align}
\end{subequations}
The last two of them contain products of oscillatory factors with frequencies
$\Delta\omega$ and $\Omega$. As a result, their time integrals involve the
combinations $\Delta\omega\pm\Omega$, which can lead to a significant enhancement when the two characteristic frequencies become comparable.
Without loss of generality, we set $y=z=0$, since these coordinates appear only through phase combination $\Delta k_y y+\Delta k_z z$, and can be absorbed into a redefinition of $\alpha$. The backflow functional is evaluated according to~(\ref{bfunc}). To keep the expressions transparent, the individual contributions are treated separately. The orbital part reads
\begin{eqnarray}\label{ocup}
F^{(o)}(T)&=&\hspace{-1ex}\int\limits_{t_0-\frac{T}{2}}^{t_0+\frac{T}{2}}\hspace{-1ex}\dd t\,j_z^{(o)}(t)=\frac{\hbar}{2m}\Big[\left(k_z^{(1)}\phi_{01}^2+k_z^{(2)}\phi_{02}^2\right)T\nonumber\\
&&\hspace{-1ex}+\,2\frac{k_z^{(1)}+k_z^{(2)}}{\Delta\omega}\phi_{01}\phi_{02}\sin\frac{\Delta\omega T}{2}\,\cos (\Delta \omega t_0-\alpha)\Big].\nonumber\\ 
\end{eqnarray}
The first term is purely static and grows linearly with the observation time $T$. The second term originates from the interference between the two modes and remains oscillatory, potentially generating negative contributions to the probability flow.

\begin{figure*}[t]
\centering
\includegraphics[width=0.90\textwidth,angle=0]{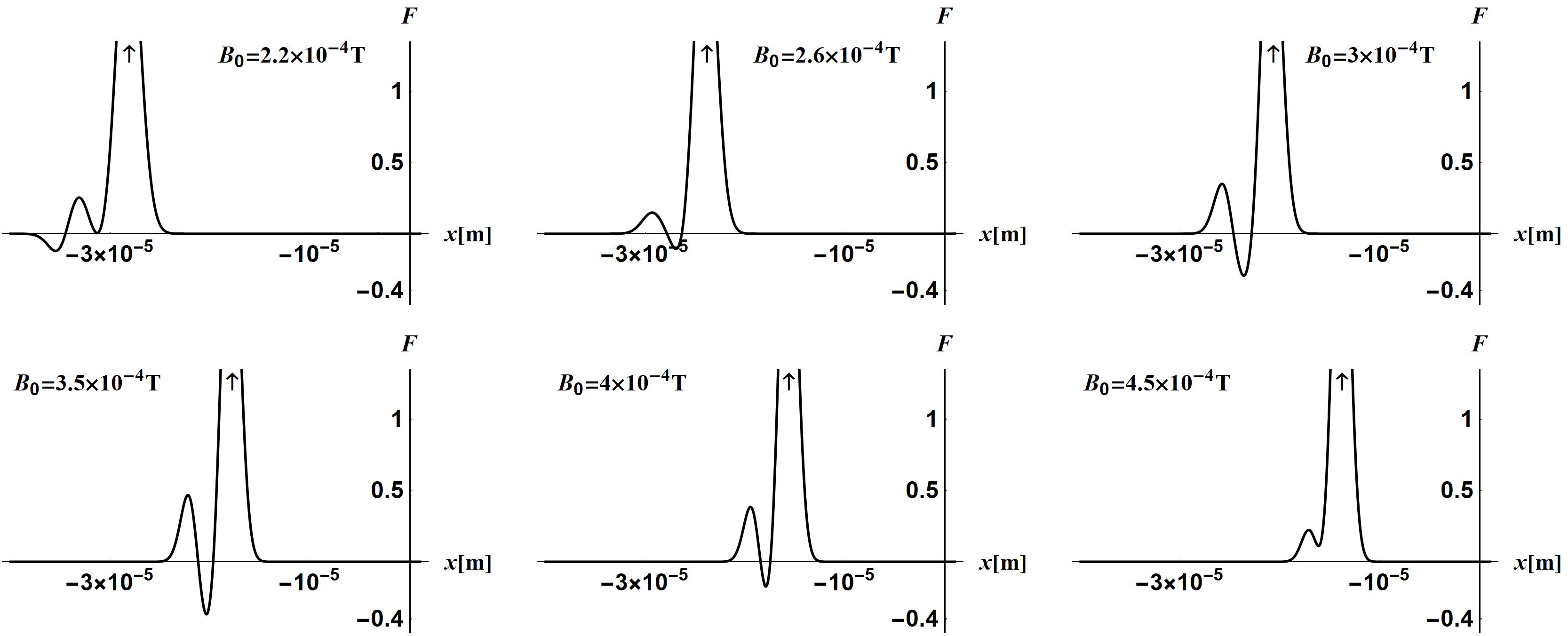}
\caption{Dependence of the backflow functional $F$ in the two-mode regime including spin contributions, shown for six selected values of the external magnetic field $B_0$. The parameters are fixed to $T=1.42\times 10^{-7}\,\mathrm{s}$, $k_y^{(1)}=10^{7}\,\mathrm{m}^{-1}$, $k_y^{(2)}=1.2\,k_y^{(1)}$, $k_z^{(1)}=0.9\times 10^{6}\,\mathrm{m}^{-1}$, and $k_z^{(2)}=0.05\,k_z^{(1)}$.}
\label{twomodes}
\end{figure*} 

The local spin contribution reflects the spatial structure of the individual Landau orbitals:
\begin{eqnarray}\label{eftsl}
F_{\mathrm{loc}}^{(s)}(T)&=&
\hspace{-1ex}\int\limits_{t_0-\frac{T}{2}}^{t_0+\frac{T}{2}}\hspace{-1ex} \dd t\,j_{z,\mathrm{loc}}^{(s)}(t)\\
&=&-\frac{\hbar}{m\Omega}\partial_x\left(\phi_{01}^2+\phi_{02}^2\right)
\sin\frac{\Omega T}{2}\sin(\Omega t_0-\chi).\nonumber
\end{eqnarray}
It contains no inter-mode interference and does not depend on $\Delta k_y$ or $\Delta k_z$, acting as a time-dependent modulation of the background through the density gradient.

The gradient term couples the orbital overlap to the transverse interference pattern:
\begin{eqnarray}\label{eftslg}
F_{\mathrm{grad}}^{(s)}(T)&=&\hspace{-1ex}\int\limits_{t_0-\frac{T}{2}}^{t_0+\frac{T}{2}}\hspace{-1ex}\dd t\,j_{z,\mathrm{grad}}^{(s)}(t)=\frac{\hbar}{2m}\partial_x\left(\phi_{01}\phi_{02}\right)\\
&&\hspace{-1ex}\times\bigg[\frac{\sin\frac{(\Delta\omega-\Omega)T}{2}}{\Delta\omega-\Omega}\,\sin\left((\Delta\omega-\Omega)t_0+\alpha-\chi\right)\nonumber\\
&&\hspace{-1ex}-\frac{\sin\frac{(\Delta\omega+\Omega)T}{2}}{\Delta\omega+\Omega}\,\sin\left((\Delta\omega+\Omega)t_0-\alpha-\chi\right)\bigg].\nonumber
\end{eqnarray}
while the interference term represents a direct spin-induced modulation of the longitudinal current:
\begin{eqnarray}\label{eftsli}
F_{\mathrm{int}}^{(s)}(T)&=&\hspace{-1ex}\int\limits_{t_0-\frac{T}{2}}^{t_0+\frac{T}{2}}\hspace{-1ex}\dd t\,j_{z,\mathrm{int}}^{(s)}(t)=-\frac{\hbar\Delta k_y}{2m}\,\phi_{01}\phi_{02}\\
&&\hspace{-1ex}\times\bigg[\frac{\sin\frac{(\Delta\omega-\Omega)T}{2}}{\Delta\omega-\Omega}\,\sin\left((\Delta\omega-\Omega)t_0+\alpha-\chi\right)\nonumber\\
&&\hspace{-1ex}+\frac{\sin\frac{(\Delta\omega+\Omega)T}{2}}{\Delta\omega+\Omega}\,\sin\left((\Delta\omega+\Omega)t_0-\alpha-\chi\right)\bigg].\nonumber
\end{eqnarray}
A noteworthy feature of Eqs.~(\ref{eftslg}) and (\ref{eftsli}) is the appearance
of the frequency combinations $\Delta\omega\pm\Omega$ in the denominators. When the spin-precession frequency approaches the orbital frequency difference, i.e.
$\Omega\simeq\Delta\omega$, the corresponding contribution scales as
$\sin[(\Delta\omega-\Omega)T/2]/(\Delta\omega-\Omega)\approx T/2$.
Consequently, the spin-dependent part of the backflow is significantly enhanced. 
The strength of this enhancement can be further tuned through the transverse
momentum mismatch $\Delta k_y$, which enters directly in
Eq.~(\ref{eftsli}). Away from resonance, for $|\Delta\omega-\Omega|T\gtrsim 1$, the full oscillatory dependence is retained.

We now fix the phase parameters $\alpha$ and $\chi$ to simplify the subsequent analysis and make the interference structure more transparent. As discussed above, $\alpha$ effectively combines the relative phase of the two modes with the choice of observation point in the $yz$ plane. The observation time $t_0$ is treated in the same spirit and is chosen to enhance the interference contribution.

We focus on the range $\Omega \approx \Delta\omega$, where the spin and orbital time scales become comparable (the case $\Omega \approx -\Delta\omega$ can be treated in an analogous manner). It is then convenient to impose $\alpha-\chi=\pi/2$ and to choose
$t_0=\pi/(2\Delta\omega)$, since at $\Omega=\Delta\omega$ this gives
\begin{equation}\label{warsi}
\sin\left((\Delta\omega-\Omega)t_0+\alpha-\chi\right)=1.
\end{equation}
A convenient choice is $\alpha=\pi/2$ and $\chi=0$.

The full backflow functional is given by
\begin{equation}
F(T)=F^{(o)}(T)+F_{\mathrm{loc}}^{(s)}(T)+F_{\mathrm{grad}}^{(s)}(T)+F_{\mathrm{int}}^{(s)}(T),
\end{equation}
and with the above choice of parameter values the orbital contribution becomes
\begin{eqnarray}\label{ostfo}
F^{(o)}(T)&=&\frac{\hbar}{2m}\bigg[\left(k_z^{(1)}\phi_{01}^2+k_z^{(2)}\phi_{02}^2\right)T\\
&&+\frac{2\left(k_z^{(1)}+k_z^{(2)}\right)}{\Delta\omega}\,\phi_{01}\phi_{02}\sin\frac{\Delta \omega T}{2}\bigg],
\nonumber
\end{eqnarray}
and the spin contributions read
\begin{subequations}\label{ostf}
\begin{align}
&F^{(s)}_{\mathrm{loc}}(T)=2\operatorname{sgn}(\Omega)\big[(x-x_1)\phi_{01}^2+(x-x_2)\phi_{02}^2\big]\nonumber\\
&\hspace{10ex}\times\sin\frac{\Omega T}{2}\sin\frac{\pi\Omega}{2\Delta \omega},\label{ostfsl}\\[0.3em]
&F^{(s)}_{\mathrm{grad}}(T)=-|\Omega|(x-x_c)\phi_{01}\phi_{02}\sin\frac{\pi\Omega}{2\Delta \omega}\label{ostfsg}\\
&\hspace{11ex}\times\bigg[\frac{\sin\frac{(\Delta\omega-\Omega)T}{2}}{\Delta\omega-\Omega}-\frac{\sin\frac{(\Delta\omega+\Omega)T}{2}}{\Delta\omega+\Omega}\bigg],\nonumber\\[0.3em]
&F^{(s)}_{\mathrm{int}}(T)=-\frac{\hbar\Delta k_y}{2m}\phi_{01}\phi_{02}\sin\frac{\pi\Omega}{2\Delta \omega}\label{ostfsi}\\
&\hspace{11ex}\times\bigg[\frac{\sin\frac{(\Delta\omega-\Omega)T}{2}}{\Delta\omega-\Omega}+\frac{\sin\frac{(\Delta\omega+\Omega)T}{2}}{\Delta\omega+\Omega}\bigg].
\nonumber
\end{align}
\end{subequations}

Figure~\ref{twomodes} shows the spatial dependence of the backflow functional for fixed $T$ and several representative values of the magnetic field $B_0$. A clear enhancement of the signal is observed near $B_0 \simeq 3\times 10^{-4}\,\mathrm{T}$. For the chosen parameters, the resonance condition $\Omega \simeq \Delta\omega$ predicts $B_0^{\mathrm{res}} \simeq 2.5\times 10^{-4}\,\mathrm{T}$. The small shift of the maximum is due to non-resonant contributions retained in the full expression for $F(T)$. The particular parameter values are not generic, since interference phenomena involving the interplay of spatial and spin degrees of freedom are inherently sensitive to the relative values of the parameters.

\begin{figure}[h!]
\centering
\includegraphics[width=0.37\textwidth]{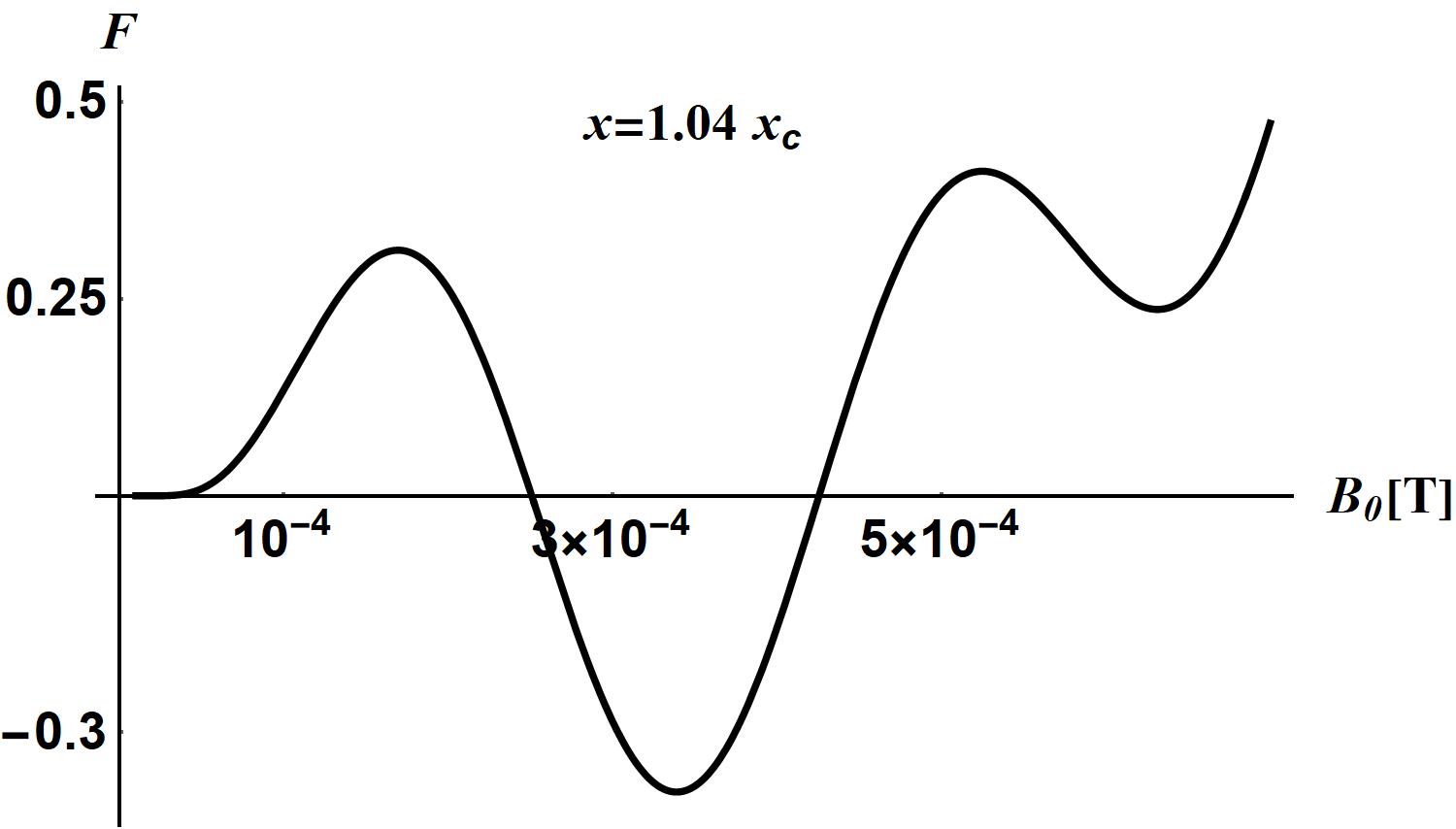}
\caption{Dependence of the backflow functional on the magnetic field strength $B_0$ at a fixed spatial point $x=1.04\,x_c$. Other parameters are the same as in Fig. \ref{twomodes}.}
\label{bdep}
\end{figure}

Figure~\ref{bdep} shows the corresponding resonant backflow profile as a function of $B_0$ at a fixed position $x=1.04\,x_c$. The same parameters are used as in Fig.~\ref{twomodes}. The agreement between both plots confirms that the enhancement is not a purely local feature in space, but reflects a global interference mechanism controlled by the magnetic field.

These results demonstrate that backflow in the spinful two-mode Landau system can be efficiently controlled by the magnetic field. It simultaneously sets the spatial separation of the guiding centers and the spin-precession time scale, and their interplay leads to an enhanced interference contribution to the probability current.

\subsection{Finite resolution and robustness of the backflow}\label{res}

In any realistic measurement, the probability current is accessed with finite spatial resolution rather than at a single point. The experimentally relevant quantity is therefore a coarse-grained functional
\begin{equation}\label{ftxy}
\overline{F}(T;x,y)=\int dx' dy'\,
W_{\delta x,\delta y}(x-x',y-y')\,F(T;x',y'),
\end{equation}
where $W_{\delta x,\delta y}$ is a normalized detector response function with characteristic widths $\delta x$ and $\delta y$.

The transverse structure of the problem discussed above is governed by two intrinsic length scales. In the $x$ direction, the magnetic field introduces the characteristic length $\ell_B=\sqrt{\hbar/|qB_0|}$, which controls both the width of individual Landau orbitals and the extent of their overlap. In the $y$ direction, interference is set by the phase factor $\exp(i\Delta k_y y)$, giving the independent scale $1/|\Delta k_y|$.

Expanding around the optimal observation point $(x_0,y_0)$, where linear variations vanish, one obtains
\begin{eqnarray}\label{fcrd}
F(T;x,y)&=&F(T;x_0,y_0)\\
&&+\mathcal{O}\!\left(\frac{(x-x_0)^2}{\ell_B^2},(\Delta k_y (y-y_0))^2\right).\nonumber
\end{eqnarray}
For symmetric detector profiles, the leading corrections are quadratic, and therefore
\begin{equation}\label{ftprz}
\overline{F}(T;x_0,y_0)=F(T;x_0,y_0)+
\mathcal{O}\!\left(\frac{\delta x^2}{\ell_B^2},(\Delta k_y\,\delta y)^2\right).
\end{equation}

Equation (\ref{ftprz}) shows the key point: backflow is stable under finite spatial resolution as long as the detector does not exceed the intrinsic scales of the system, i.e.,
\begin{equation}\label{parpr}
\delta x \lesssim \ell_B,\qquad \delta y \lesssim \frac{1}{|\Delta k_y|}.
\end{equation}

The relation $|\Delta x|=\ell_B^2|\Delta k_y|$ connects orbital separation with transverse interference. Appreciable overlap, $|\Delta x|\lesssim \ell_B$, therefore requires $|\Delta k_y|\lesssim 1/\ell_B$.  This identifies the crossover regime in which interference and Landau localization are simultaneously relevant and the backflow signal is maximally robust.

For the parameters of Figs.~\ref{twomodes} and \ref{bdep}, $\ell_B\approx 1.5\times10^{-6}\,\mathrm{m}$ and $1/|\Delta k_y|\sim5\times10^{-7}\,\mathrm{m}$, placing the system in the crossover range. These scales are well within current cold-atom and optical imaging capabilities, implying that the predicted backflow enhancement is experimentally accessible and not an artifact of point-like detection.

An analogous consideration applies to temporal resolution. Averaging over a finite time window $\delta t$ limits the resolution of interference oscillations governed by $1/\Delta\omega$ and $1/\Omega$. Observable backflow therefore requires $\delta t \lesssim 1/\max(\Delta\omega,\Omega)$. For the parameters of Figs.~\ref{twomodes} and \ref{bdep}, with $B_0\sim3\times10^{-4}\,\mathrm{T}$, the characteristic time scales near resonance, $\Omega\approx\Delta\omega$, are $10^{-8}\text{--}10^{-7}\,\mathrm{s}$, corresponding to experimentally accessible resolutions in the tens of nanoseconds regime.

\section{Summary and conclusions}\label{sum}

Probability backflow in a uniform magnetic field has been analyzed within the Schr\"odinger and Pauli frameworks, with emphasis on magnetic-field control. For a single Landau orbital, the Schr\"odinger current remains positive for positive longitudinal momentum, precluding backflow. In the Pauli case, the spin contribution can locally produce negative current through its oscillatory dynamics. However, the effect remains weak and confined by the Gaussian Landau envelope, preventing a robust backflow structure.

A qualitatively different behavior is obtained for two-mode superpositions, where backflow arises from interference between states with different longitudinal momenta. The magnetic field then provides a control parameter by governing both the spatial overlap of the guiding centers and the spin-precession dynamics, thereby controlling the strength and spatial profile of the backflow.

In the Pauli two-mode system, a pronounced enhancement is obtained when the spin-precession frequency approaches the orbital frequency mismatch. The resulting resonance amplifies the interference contribution and allows the backflow to be tuned by the magnetic field.

Finite spatial resolution does not destroy the effect, provided that the detector resolution remains smaller than both the magnetic length and the interference scale. The predicted backflow is therefore robust against coarse-graining rather than being an artifact of point-like detection.

Overall, backflow in Landau systems is an interference-driven effect whose strength can be controlled by the magnetic field, with spin playing an essential role in the two-mode regime. Possible extensions include higher Landau levels, multimode configurations, and weak magnetic-field inhomogeneities.

\end{document}